\documentclass[prl,twocolumn,a4paper,superscriptaddress,showpacs]{revtex4}
\usepackage{graphicx}

\newcommand{\be}{\begin{equation}}
\newcommand{\ee}{  \end{equation}}
\newcommand{\ba}{\begin{eqnarray}}
\newcommand{\ea}{  \end{eqnarray}}

\begin{document}

\title{Coherent versus sequential electron tunneling in quantum dots}

\author{L. E. F. Foa Torres}
\affiliation{Facultad de Matem\'{a}tica, Astronom\'{\i}a y
             F\'{\i}sica, Universidad Nacional de C\'{o}rdoba,\\ 
             Ciudad Universitaria, 5000 C\'{o}rdoba, Argentina}

\author{C. H. Lewenkopf}
\affiliation{Instituto de F\'{\i}sica, 
             Universidade do Estado do Rio de Janeiro, \\  
             R. S\~ao Francisco Xavier 524, 20550-900 Rio de Janeiro, Brazil}

\author{H. M. Pastawski}
\affiliation{Facultad de Matem\'{a}tica, Astronom\'{\i}a y
             F\'{\i}sica, Universidad Nacional de C\'{o}rdoba,\\ 
             Ciudad Universitaria, 5000 C\'{o}rdoba, Argentina}

%%%%%%%%%%%%%%%%%%%%%%%%%%%%%%%%%%%%%%%%%%%%%%%%%%%%%%%%%%%%%%%%%%%%%%%%%%
%                            ABSTRACT                                    %
%%%%%%%%%%%%%%%%%%%%%%%%%%%%%%%%%%%%%%%%%%%%%%%%%%%%%%%%%%%%%%%%%%%%%%%%%%
\begin{abstract}
Manifestations of quantum coherence in the electronic conductance through 
nearly closed quantum dots in the Coulomb blockade regime are addressed. 
We show that quantum coherent tunneling processes explain some puzzling 
statistical features of the conductance peak-heights observed in recent 
experiments at low temperatures.
We employ the constant interaction model and the random matrix theory to model
the quantum dot electronic interactions and its single-particle statistical
fluctuations, taking full account of the finite decay width of the quantum
dot levels.

\end{abstract}

%
% 03.65.Yz  Decoherence; open systems; quantum statistical methods 
% 73.21.La  Quantum dots 
% 73.23.-b  Mesoscopic systems 
% 73.23.Ad  Ballistic transport 

\maketitle

Recent experimental studies of electronic transport through nearly
isolated quantum dots \cite{Patel98,Folk01} assess the importance of 
quantum coherence and the nature of dephasing mechanisms in finite 
interacting electronic systems.
Of particular interest is the Coulomb blockade regime, where the thermal 
energy $k_BT$ is much smaller than the charging energy $E_C$ necessary 
to add an electron to the quantum dot.
In this regime the conductance depends primarily on the quantum 
properties of the dot, such as its resonance levels and the corresponding 
line widths due to the coupling between the dot and leads.
Electrons are allowed to tunnel through the quantum dot whenever the 
charging energy is compensated by an external potential and the dot energy levels
are in resonance with the chemical potential at the leads (small bias limit). 
The tunneling condition can be attained, for instance, by a tunable gate 
voltage $V_g$. In a typical experiment $V_g$ is varied to obtain the
conductance spectrum, a sequence of sharp (Coulomb blockade) peaks.

Sequential tunneling is the key hypothesis for the standard rate equations 
\cite{Beenakker91} used to explain the transmission spectrum of quantum 
dots in the Coulomb blockade regime \cite{Alhassid00,Aleiner02}. 
This probabilistic picture neglects non-resonant quantum virtual processes, 
under the assumption that the resonant decay widths $\Gamma $ are
much smaller than both $k_{B}T$ and the energy separation between the quantum 
dot resonances $\delta \varepsilon $, namely, $\Gamma \ll k_{B}T$ and $\,\Gamma 
\ll \delta \varepsilon $, a condition often met by experiments in nearly isolated 
quantum dots. 

The early experimental data taken from ballistic chaotic quantum dots were
successfully confronted with the sequential theory by using the
random matrix theory (RMT) to model the dot statistical single-particle 
properties \cite{Alhassid00,Aleiner02}. 
More recently, the analysis of the measured conductance peak-heights in the 
Coulomb blockade regime \cite{Patel98,Folk01} show significant deviations 
from this theory \cite{Jalabert92,Alhassid98,Vallejos99}, indicating that
some physics is missing.
The inclusion of inelastic scattering processes \cite{Beenakker01,Eisenberg02,Held02,Rupp02}, 
spin-orbit coupling \cite{Held02-spin-orbit}, and exchange interaction 
\cite{Usaj02,Alhassid02} into the sequential approach expand in interesting ways
the considered physical processes, adding new parameters to the description. Unfortunately, these studies achieved only a limited  success in reconciling theory with experiment. 

In this Letter we show that quantum coherence, so far overlooked, leads to
important corrections to the sequential tunneling picture \cite{Weil87} and explains some 
of the puzzles pointed out by the conductance experiments \cite{Patel98,Folk01}.
The importance of coherent processes is justified by
noticing that while the sequential theory requires $\Gamma \ll k_{B}T,$ $%
\delta \varepsilon ,$ the experiments satisfy those conditions only in
average, namely, $\langle \Gamma \rangle <\Delta \equiv \langle \delta
\varepsilon \rangle$ and $\langle \Gamma \rangle \lesssim k_{B}T$. Since
both the decay width $\Gamma $ and the resonance spacings $\delta
\varepsilon $ fluctuate, conductance peaks where $\Gamma $ is larger than $%
k_{B}T$ and comparable to $\delta \varepsilon $ are not exceptional. 
%In this work our aim
%is to address the effect of \ non negligible coupling to the leads on the
%distribution of conductance peak heights. We show that this can lead to
%significant deviations from the sequential theory at low temperatures.
More importantly, the study of fully coherent transport, as opposed
to the sequential tunneling limit, provides a better framework 
to understand the interplay between coherence and interactions.

We describe a quantum dot coupled to external leads by the Hamiltonian 
\begin{equation}
\hat{H}=\hat{H}_{\mathrm{dot}}+\hat{H}_{\mathrm{leads}}+\hat{H}_{\mathrm{%
coupling}}.
\end{equation}
We write the chaotic quantum dot Hamiltonian $\hat{H}_{\mathrm{dot}}$ as 
\begin{equation}  \label{eq:Hdot}
\hat{H}_{\mathrm{dot}}=\sum_{j}(E_{j}-e\eta V_{g})d_{j}^{+}d_{j}^{{}}+ 
\frac{e^{2}}{2C}\hat{N}\left(\hat{N}-1\right) ,
\end{equation}
where $d_{j}^{+}$ creates an electron in the $j$th eigenstate with energy $%
E_j$ of the closed dot, $\hat{N}=\sum_{j}d_{j}^{+}d_{j}^{{}}$ is the
electron number operator in the dot, $\eta V_{g}$ is the electrostatic
energy due to the external gate (as usual, $V_{g}$ is the gate voltage and $\eta$
depends on the system specifics), and $C$ is the effective dot capacitance.
Equation\ (\ref{eq:Hdot}) is the constant interaction model. In chaotic quantum
dots ground state energy fluctuations due to interaction effects are very
small in the large $N$ limit \cite{Aleiner02}. We also do not account for
spin and exchange interaction, which were recently addressed in the master
equation framework by Refs. \cite{Usaj02,Alhassid02}. The electrons in the
leads are treated as non-interacting, namely 
\begin{equation}
\hat{H}_{\mathrm{leads}}=\sum_{k,\mathrm{a}\in \mathrm{L,R}}\varepsilon_{k,%
\mathrm{a}}^{} c_{k,\mathrm{a}}^{+}c_{k,\mathrm{a}}^{{}}\,,
\end{equation}
where $c_{k,\mathrm{a}}^{+}$ creates an electron at the state of wave vector 
$k = (2m^*\varepsilon_k)^{1/2}/\hbar$ at channel ``a'' either in the left
(L) or in the right (R) lead. The dot-lead coupling term is 
\begin{equation}
\hat{H}_{\mathrm{coupling}}=\sum_{k,\mathrm{a}\in \mathrm{L,R}}\sum_{j}
\left(V_{(k,\mathrm{a}),j}^{{}}c_{k,\mathrm{a}}^{+}d_{j}^{{}}+h.c.\right) \,.
\end{equation}
The magnitude of the coupling matrix elements $V_{(k,\mathrm{a}),j}^{}$ 
determine through a Fermi golden rule \cite{cit-HoracioGLBE2} the electron 
decay width $\Gamma$, or the tunneling rate $\Gamma/\hbar$ in
the master equation framework. For quantum dots in the Coulomb blockade regime $%
\langle \Gamma \rangle$ is much smaller than the dot mean level spacing $%
\Delta$.

The conductance through the quantum dot is expressed in terms of the
interacting system retarded Green's function, $G_{i,j}^{R}(t) =-(\mathrm{i}%
/\hbar)%
\Theta (t)\left\langle \left\{d_{i}^{{}}(t),d_{j}^{+}(0) \right\}
\right\rangle$. The evaluation of $G_{i,j}^{R}(t)$ follows the treatment
presented by Baltin and collaborators \cite{Baltin99} and generalizes their
result to cases where the condition $\Gamma \ll \delta \varepsilon$ is not
met. 

The retarded Green's function is written as a sum over terms
containing different (and fixed) number of electrons in the dot 
\begin{equation}  \label{eq:GR}
G_{i,j}^{R}(t)= -\frac{\mathrm{i}}{\hbar}\Theta (t)\sum_{N=0}^\infty
P_{N}\left\langle \left\{d_{i}^{{}}(t),d_{j}^{+}(0)\right\}
\right\rangle _{N} \,,
\end{equation}
where $P_{N}$ is the thermal probability to find $N$ electrons in the dot.
This probability considers the full set of occupation numbers 
$\{n_\ell\}$ of the $\hat{H}_{\mathrm{dot}}$ eigenstates. 
Equation (\ref{eq:GR}) can be formally solved by the
method of equation of motion. In practice, the equations do not close unless
we assume that the number of electrons in the dot does not fluctuate, which
means that we replace $\hat{N}$ by its expectation value $N$ \cite{Meir91}.
This simplification is entirely justified in the cases of interest, where $%
e^2/C \gg \mathrm{max}\,(\Gamma, k_B T)$.

The matrix representation of the retarded Green's function is then casted as 
\begin{eqnarray} \label{eq:GR-energy}
G^{R}=\sum_{N=0}^{\infty }P_{N} &&\!\!\!\!\!\!\!\Big\{\left[ \varepsilon
I-H_{\mathrm{dot}}^{(N)}-\Sigma ^{R}(\varepsilon )\right] ^{-1}(I-n_{N})+ 
\nonumber  \label{eq:GRexplicit} \\
&&\!\!\left[ \varepsilon I-H_{\mathrm{dot}}^{(N-1)}-\Sigma ^{R}(\varepsilon )%
\right] ^{-1}n_{N}\Big\}.
\end{eqnarray}
where the quantum dot matrix elements are 
\begin{equation}
\left[ H_{\mathrm{dot}}^{(N)}\right] _{i,j}=(E_{j}-e\eta V_{g}+UN)\delta
_{i,j},
\end{equation}
and $U$ is the quantum dot charging energy, namely, $U=e^{2}/C$. In Eq.\ (%
\ref{eq:GRexplicit}) we define $\left[ n_{N}\right] _{i,j}=\left\langle
n_{i}\right\rangle _{N}\delta _{i,j}$ as the diagonal matrix whose entries
are the canonical occupation numbers of the (closed) dot eigenstates. The retarded
self-energy matrix elements, due to the coupling to the leads, become 
\begin{equation}
\left[ \Sigma ^{R}(\varepsilon )\right] _{i,j}=\sum_{k,\mathrm{a}\in \mathrm{%
L,R}}\frac{V_{i,(k,\mathrm{a})}^{{}}V_{(k,\mathrm{a}),j}^{{}}}{\varepsilon +%
\mathrm{i}0^{+}-\varepsilon _{k,\mathrm{a}}}\,.
\end{equation}
The coupling matrix elements $V_{(k,\mathrm{a}),j}^{{}}$ vary in the energy
scale of $\varepsilon _{k}$ and hence are practically constant in energy
windows comprising several single-particle states. We neglect such
variations to write 
\begin{equation}
\Sigma ^{R}(\varepsilon )=-\frac{\mathrm{i}}{2}\left( \Gamma _{\mathrm{L}%
}+\Gamma _{\mathrm{R}}\right)
\end{equation}
where $\sum_{k}{V_{i,(k,\mathrm{a})}^{}V_{(k,\mathrm{a}),j}^{{}}}%
/(\varepsilon +\mathrm{i}0^{+}-\varepsilon _{k,\mathrm{a}})=-\mathrm{i}\left[
\Gamma _{\mathrm{a}}\right] _{i,j}/2$. The energy dependence due to the
principal value integral is also negligible in the Coulomb blockade regime,
since there are no open transmitting channels.

The linear-response conductance is \cite{cit-HoracioGLBE2} 
\begin{equation}
G=\frac{e^{2}}{h}g\quad \mathrm{with}\quad g=\int \!d\varepsilon \!\left( -%
\frac{\partial f_{\mu }}{\partial \varepsilon }\right) T_{\mathrm{R,L}%
}(\varepsilon )\,,
\end{equation}
where $f_{\mu }$ is the Fermi distribution function in the leads with
chemical potential $\mu $. $T_{R,L}$ is the system transmittance that can be
directly computed from the retarded Green's function 
\begin{equation}
T_{R,L}(\varepsilon )=\Big|\sum_{i,j}V_{(k,\mathrm{L}),i}^{{}}\left[ G^{R}%
\right] _{i,j}V_{j,(k,\mathrm{R})}^{{}}\Big|^{2}\,.  \label{eq:T_RL}
\end{equation}
Equivalently, the above expression can also be casted in the well-known form 
$T_{R,L}=\mathrm{tr}\,(\Gamma _{\mathrm{R}}G^{R}\Gamma _{\mathrm{L}}G^{A})$ 
\cite{cit-HoracioGLBE2}.

To this point our approach is quite general. The only important
approximation we make requires $e^{2}/C\gg \mathrm{max}\,(\Gamma ,k_{B}T)$.
Albeit restrictive, the approximation is compatible with the Coulomb 
blockade experiments we are interested in. 
Our approach is reduced to the sequential
tunneling one \cite{Beenakker91} in the limit of $\Gamma \ll \mathrm{min}%
\,(k_{B}T,\delta \varepsilon )$. The main improvement is that we naturally
account for quantum virtual tunneling processes. Those are significant 
whenever $k_{B}T$ becomes comparable with $\Gamma $, a condition
often met by experiments. Furthermore, both the single-particle level
spacings $\delta \varepsilon $ and the decay widths $\Gamma $ fluctuate.
Even if in average $\Delta \gg \langle \Gamma \rangle $, situations where $%
\delta \varepsilon $ is comparable to $\Gamma $ are not infrequent. In these
cases quantum corrections are important. 
When the condition $\Gamma /\delta \varepsilon \ll 1$ is always satisfied and 
\textsl{not only in average}, corrections to
the conductance become indeed negligible. 
This was the limit analyzed in Ref. \cite{Baltin99} for the phase lapse problem.
Note also the contrast with the case of elastic cotunneling at the conductance 
valleys. There, the contribution of the off-resonant levels is of order 
$\Gamma /U$, whereas here their contribution is of order $\Gamma /\delta 
\varepsilon$.

We switch now to the statistical study of the dimensionless conductance peak
heights $g^{\mathrm{max}}$. This analysis allows for a comparison between the 
results of
our approach, experiments and the sequential tunneling theory. The
statistical ansatz is to assume that the underlying electronic dynamics in
the quantum dot is very complex and hence the fluctuation properties of
its single-particle eigenenergies and eigenfunctions coincide with those of an
ensemble of random matrices \cite{Alhassid00,Aleiner02}. Accordingly, the
single-particle levels display universal fluctuations and their spacings $%
\delta \varepsilon$ follow the Wigner-Dyson distribution. Likewise, the
decay widths $\Gamma$ are Porter-Thomas distributed. The  inputs of
the statistical theory are  the mean level spacing $\Delta$ and the
average decay width $\langle \Gamma \rangle$. We consider the dot both in the
absence of a magnetic field (orthogonal ensemble, $\beta=1$) and in the presence 
of a magnetic field $B$ that breaks the time-reversal-symmetry (unitary ensemble, $\beta=2$). 
The later is the relevant one for comparison with avaliable experimental data.

The numerical implementation is straightforward, but costly since 
Eq.\ (\ref{eq:GR-energy}) requires matrix inversions for each realization.
The canonical thermal quantities $P_{N}$ and $\left\langle
n_{i}\right\rangle_{N}$ are computed using the quadrature formula explained
in Ref. \cite{cit-Ormand}, already used for quantum dots 
\cite{Alhassid98,Vallejos99}. For $k_BT \lesssim \Delta$ good individual peak 
height accuracy requires taking into account at least 30 levels around the 
resonant one. 
Between $5\times10^{4}$ and $1\times10^{5}$ realizations were used for the 
ensemble averaging.
The charging energy $U$ is taken to be $50\Delta$ (the results are quite 
insensitive to $U $, provided $U \gg \Delta$).

\begin{figure}[tbp]
%\vspace{0.5cm} 
%\centering \leavevmode
%\center{\epsfig{file=Pg.eps, ,width=8.0cm,angle=0}}
\includegraphics[width=8.0cm]{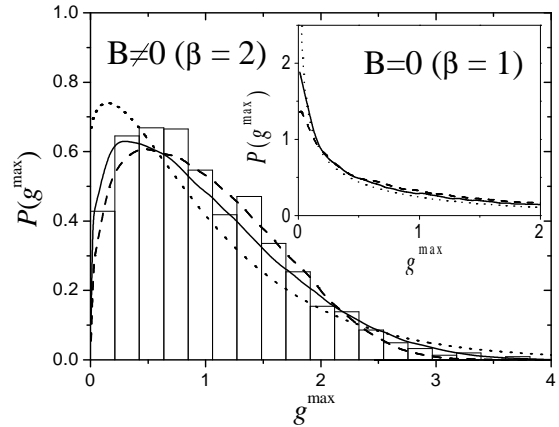} \vspace{-0.3cm}
\caption{Peak height probability distribution $P(g^{\mathrm{max}})$ for $%
k_{B}T=0.1\Delta$ and $B\neq0$ ($\beta=2$). 
The same for $B=0$ ($\beta = 1$) in the inset.
Our theory for $\langle\Gamma\rangle/\Delta =0.1$ (solid line) and 0.2 
(dashed line) is compared with the standard sequential tunneling result 
(dotted line), and the experimental distribution (histogram) 
\protect\cite{Patel98}. }
\label{fig-Pg}
\end{figure}

The data of Ref. \cite{Patel98} show that at very low temperatures, $%
k_{B}T\ll \Delta $, the conductance peak-height distribution does not follow
the standard random matrix theory \cite{Jalabert92}.  
By accounting for quantum coherent tunneling we obtain a very nice agreement
with the experimental distributions.
This is illustrated in Fig. \ref{fig-Pg} for $B\neq0$ ($\beta =2$). 
In the inset we present our results for the distribution of $g^{\max}$ for 
$B=0$ ($\beta=1$). 
In Fig. \ref{fig-Pg} the dimensionless conductance peak heights $g^{\max }$ 
are scaled to unit mean. 
We show the peak heights distribution for $k_{B}T=0.1\Delta $, 
$\left\langle \Gamma \right\rangle =0.1\Delta $ (solid line)\ and 
$\left\langle \Gamma \right\rangle =0.2\Delta $ (dashed line).
The histogram corresponds to the experimental result of Ref. \cite{Patel98}
available only for $B\neq0$ ($\beta=2$).
Different dots have different $\langle \Gamma \rangle/\Delta$, a ratio that
can be determined from the experimental $g^{\max}$.
$\langle \Gamma \rangle/\Delta \sim 0.1$ is representative of the analyzed
experiments.
We find that as the ratio $\left\langle \Gamma \right\rangle /\Delta $ is
increased, the probability to obtain small conductances is suppressed in
comparison with the standard sequential theory (dotted line). This can be
understood as follows: If a given resonance has small tunneling rates, the
contributions due to virtual processes through its neighbors will reduce 
the chance to obtain a very small peak. Thus, we expect $P(g^{\max}=0)=0$. 

In the early experiment by Chang \textit{et al.}\ \cite{Chang96} special care
was taken to discard from the statistical sample conductance peak-heights that
did not fulfill $\Gamma \ll k_{B}T$. Hence, corrections due to the finite ratio 
$\Gamma /\Delta $ are practically negligible. This
might explain why a good agreement with the standard sequential theory was
found there \cite{Chang96}.
Note also that as $k_{B}T$ becomes comparable with $\langle\Gamma\rangle$ 
the assessment of the quantum dot temperature through the widths of the 
Coulomb-blockade peaks becomes unreliable, due to the non-negligible $\Gamma$.

\begin{figure}[tbp]
%\vspace{0.7cm}
%\centering \leavevmode
%\center{\epsfig{file=sigmag.eps, ,width=8.0cm,angle=0}}
\includegraphics[width=8.0cm]{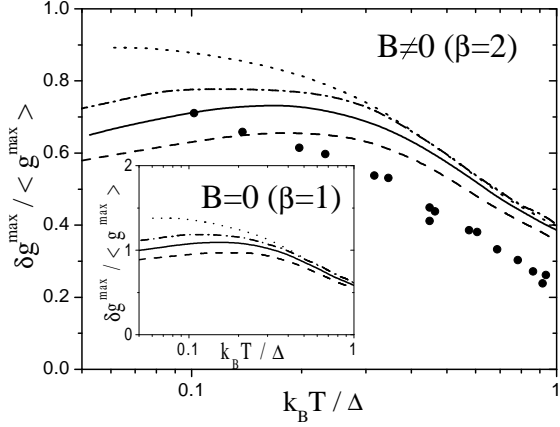} \vspace{-0.3cm}
\caption{Normalized peak heights distribution width $\protect\sigma_g$ for
$B\neq0$ (the $B=0$ case is shown in the inset) as a function 
of $k_BT/\Delta$, for $\langle\Gamma\rangle/
\Delta=0.05, 0.1, 0.2$ (dashed-dot, solid and dashed lines respectively).
Symbols correspond to the experimental results of Ref. \protect\cite{Patel98}
{\sl for different dots} and the dotted lines to the standard sequential theory.}
\label{fig-sigmag}
\end{figure}

The experimental results of Ref. \cite{Patel98} show another striking and
unexplained discrepancy with respect to the standard rate equations. This 
is best quantified by the ratio between the standard deviation $\delta g^{\max }$
and the mean conductance peak heights $\langle g^{\max }\rangle $, namely 
\begin{equation}
\sigma _{g}=\frac{\delta g^{\max}}{\left\langle g^{\max }\right\rangle }=%
\frac{\sqrt{\big\langle(g^{\max })^{2}\big\rangle-\left\langle g^{\max
}\right\rangle ^{2}}}{\left\langle g^{\max }\right\rangle }.
\label{eq-sigmag}
\end{equation}
In the experiments $\delta g^{\max }$ is significantly smaller than
predicted by the rate equations plus RMT. 
Recent works \cite{Held02,Beenakker01,Rupp02} discuss if such deviations 
can be attributed to inelastic processes \cite{comment-Patel98}. 
Our approach explains the experimental findings in the low temperature regime 
$k_BT/\Delta \ll 1$, where inelastic processes are hard to justify.  
In Fig.\ \ref{fig-sigmag} we show $\sigma _{g}$ for $B\neq0$ ($\beta =2$)
as a function of the thermal energy for different values of $\langle \Gamma \rangle
/\Delta$. 
The inset shows $\sigma _{g}$ for the case when $B=0$ ($\beta=1$).  
The standard sequential theory results \cite{Alhassid98} are illustrated by the 
dotted lines.

At low temperatures and as $\left\langle \Gamma \right\rangle /\Delta $ is increased,
our $\sigma_{g}$ is significantly reduced with respect to the standard sequential 
theory prediction.
For higher temperatures, $k_{B}T \agt 0.5\Delta $, we obtain larger $\sigma_{g}$ 
than the measured ones. Furthermore, as the temperature increases our $\sigma_{g}$ 
approaches the standard theory result. 
Similar behavior was also recently found by including the exchange term in 
$\hat{H}_{\rm dot}$ \cite{Alhassid02,Usaj02}. However, at high temperatures 
we expect a reduction of the peak heights fluctuations due to inelasticity 
and decoherence. 

\begin{figure}[tbp]
%\vspace{0.6cm} 
%\centering \leavevmode
%\center{\epsfig{file=alfa.eps, ,width=8.0cm,angle=0}}
\includegraphics[width=8.0cm]{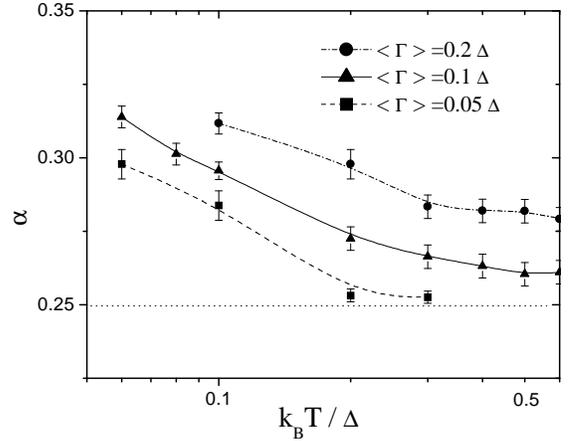} \vspace{-0.3cm}
\caption{Normalized change in the average conductance $\protect\alpha$ as a
function of temperature for different $\langle \Gamma \rangle/\Delta$.}
\label{fig-alfa}
\end{figure}

The suppression of the weak localization peak was recently used  
to determine the dephasing time $\tau_\phi$ in open quantum dots 
\cite{Huibers98,Alves02},
This inspired Folk et al. to experimentally investigate the change 
in the conductance peak-height upon breaking the time-reversal symmetry 
of the quantum dots by applying a magnetic field $B$, namely
\begin{equation}
\alpha =\frac{
   \left\langle g_{{}}^{\max }\right\rangle_{B\neq0}-
   \left\langle g_{{}}^{\max }\right\rangle_{B=0}}
  {\left\langle g_{{}}^{\max}\right\rangle_{B\neq0}}.
\end{equation}
At zero temperature the sequential tunneling theory gives a constant $\alpha = 1/4$. 
Inclusion of temperature corrections and spectral fluctuations give small 
changes, essentially keeping $\alpha \simeq 1/4$ \cite{Rupp02,Held02}.  
In Fig. \ref {fig-alfa} we show $\alpha $ as a function of temperature for different
values of $\left\langle \Gamma \right\rangle /\Delta$. 
Our simulations show
that $\alpha $ is larger than $1/4$ at low temperatures and decreases with
increasing $k_{B}T$. This behavior suggests that a finite ratio $%
\left\langle \Gamma \right\rangle /\Delta $ enhances more effectively the
conductance in the unitary case than in the orthogonal case. 
Since $\alpha $ is very sensitive to the ratio $\left\langle \Gamma
\right\rangle /\Delta $, particular care must be exercised when 
comparing data corresponding to different quantum dots. 
As in the analysis of $\sigma_g$ our results suggest that an additional
physical process is needed to explain the experimental data for 
$k_BT \agt \Delta$.

In summary, we have investigated the effect of quantum coherent processes on
the statistics of the conductance peak heights. We found that at very low
temperatures this leads to significant corrections to the distribution of
conductance peak heights obtained using the standard sequential theory. The
relevant parameter for these corrections is $\left\langle \Gamma
\right\rangle /k_BT$.
Our study also indicates that estimates of the inelastic scattering
rates and the strength of the effective exchange interaction in quantum dots 
using the peak height distributions need to account for coherent tunneling in
order to be quantitative.

L. Foa Torres and C. H. Lewenkopf thank the CBPF for the hospitality. This
work was supported by Antorchas, Vitae, CONICET, SeCyT-UNC and CNPq.


\begin{references}
\bibitem{comment-Patel98}  
  As pointed out in \cite{Beenakker01}, the word decoherence is often missused 
  in association to inelastic processes not included in the sequential tunneling theory.
\end{references}
\end{document}